\documentclass{article}

\usepackage{arxiv}

\usepackage[utf8]{inputenc} 
\usepackage[T1]{fontenc}    
\usepackage{hyperref}       
\usepackage{url}            
\usepackage{booktabs}       
\usepackage{amsfonts}       
\usepackage{nicefrac}       
\usepackage{microtype}      
\usepackage{lipsum}		
\usepackage{graphicx}
\usepackage{natbib}
\usepackage{doi}

\usepackage{rotating}
\usepackage{tablefootnote}
\usepackage{tabulary}
\usepackage{caption}
\usepackage{threeparttablex}
\usepackage{multirow}

\usepackage{pdfpages}
\usepackage{pdflscape}
\usepackage{lscape}

\usepackage{longtable}

\title{Multiple imputation for logistic regression models: incorporating an interaction}

\author{ \href{https://orcid.org/0000-0002-8502-0056}{\includegraphics[scale=0.06]{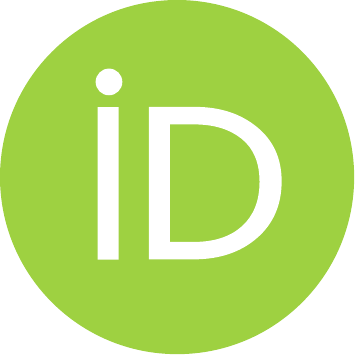}\hspace{1mm}Matthew J.~Smith} \\ 
	Inequalities in Cancer Outcomes Network\\
    Non-Communicable Disease Epidemiology\\
	London School of Hygiene and Tropical Medicine\\
	London, UK, WC1E 7HT \\
	\texttt{matt.smith@lshtm.ac.uk} \\
	\And
	\href{https://orcid.org/0000-0003-4446-0730}{\includegraphics[scale=0.06]{orcid.pdf}\hspace{1mm}Matteo Quartagno} \\
	MRC Clinical Trials Unit\\
	Institute of Clinical Trials and Methodology\\
    University College London\\
	London, UK \\
	\texttt{m.quartagno@ucl.ac.uk} \\
    \And
	\href{https://orcid.org/0000-0003-4462-2832}{\includegraphics[scale=0.06]{orcid.pdf}\hspace{1mm}Edmund Njeru Njagi} \\
    Inequalities in Cancer Outcomes Network\\
    Non-Communicable Disease Epidemiology\\
	London School of Hygiene and Tropical Medicine\\
	London, UK, WC1E 7HT \\
	\texttt{edmund.njeru.njagi@lshtm.ac.uk} \\
}

\renewcommand{\headeright}{}
\renewcommand{\undertitle}{}

\hypersetup{
pdftitle={Multiple imputation for logistic regression models: incorporating an interaction},
pdfsubject={q-bio.MJS, q-bio.MQ, q-bio.ENN},
pdfauthor={Matthew J.~Smith, Matteo Quartagno, Edmund Njeru Njagi},
pdfkeywords={Missing data, multiple imputation, interaction, simulation, substantive-model-compatible, fully-conditional-specification, socioeconomic inequalities},
}

\begin{document}
\maketitle

\begin{abstract}
    \textbf{Background} Multiple imputation is often used to reduce bias and gain efficiency
    when there is missing data. The most appropriate imputation method depends on
    the model the analyst is interested in fitting. Several imputation approaches have
    been proposed for when this model is a logistic regression model with an interaction
    term that contains a binary partially observed variable; however, it is not clear which
    performs best under certain parameter settings.
    \textbf{Methods} Using 1000 simulations, each with 10,000 observations, under six data-
    generating mechanisms (DGM), we investigate the performance of four methods:
    (i) ’passive imputation’, (ii) ’just another variable’ (JAV), (iii) ’stratify-impute-
    append’ (SIA), and (iv) ’substantive model compatible fully conditional specifica-
    tion’ (SMCFCS). The application of each method is shown in an empirical example
    using England-based cancer registry data.
    \textbf{Results} SMCFCS and SIA showed the least biased estimate of the coefficients for
    the fully, and partially, observed variable and the interaction term. SMCFCS and
    SIA showed good coverage and low relative error for all DGMs. SMCFCS had a
    large bias when there was a low prevalence of the fully observed variable in the
    interaction. SIA performed poorly when the fully observed variable in the interaction
    had a continuous underlying form.
    \textbf{Conclusion} SMCFCS and SIA give consistent estimation for logistic regression
    models with an interaction term when data are missing at random, and either can be
    used in most analyses. SMCFCS performed better than SIA when the fully observed
    variable in the interaction had an underlying continuous form. Researchers should be
    cautious when using SMCFCS when there is a low prevalence of the fully observed
    variable in the interaction.
\end{abstract}


\section{Introduction}

    Missing data is often a problem in medical research because it can cause biased and inefficient inference of parameter estimates when missing data are improperly handled, including when they are ignored (i.e., a complete case analysis), leading to reporting of incorrect conclusions. The validity of inference from incomplete data depends on the mechanism driving missingness; when used properly and assuming the correct missingness mechanism, multiple imputation is known to reduce bias and improve efficiency and is a common approach for handling missing data (\cite{Rubin1987MultipleSurveys,Carpenter2013MultipleApplication}). \\
    
    Several approaches of multiple imputation exist for modelling a continuous, binary, or time-to-event outcome along with ways of incorporating non-linear and time-varying effects or effect modification (\cite{Carpenter2013MultipleApplication}). Studies have explored scenarios that involve partially observed outcomes, and/or covariates, and interactions between partially observed covariates (\cite{VonHippel2009HowVariables,Seaman2012MultipleMethods}). However, while interactions have been extensively investigated for linear regression models, research is limited when a logistic regression model has an interaction that contains a partially observed variable (\cite{VonHippel2009HowVariables,Seaman2012MultipleMethods}). \\
    
    Medical research often focuses on regressing a binary outcome $Y$ on several covariates, one of which can be fully observed (henceforth referred as $Z$) or partially observed ($X$). For this study, we assume $Y$ is fully observed. A logistic regression model is often built with an interaction $XZ$. For example, a model built to estimate the odds of death ($Y$) amongst a cohort of cancer patients might include an interaction between the cancer diagnostic stage ($X$) and a patient's comorbidity status ($Z$). In other words, the model assumes the effect of comorbidity on mortality varies at different stages of cancer. In population-based cancer data sets, cancer diagnostic stage ($X$) is often poorly recorded, leading to studies requiring a suitable approach for multiple imputation. \\
    
    Multiple imputation methods have been suggested for logistic regression models with an interaction containing a partially observed covariate (\cite{VonHippel2009HowVariables,Seaman2012MultipleMethods}). We consider four commonly considered approaches: passive imputation, just another variable (JAV), stratify-impute-append (SIA), and substantive model compatible fully conditional specification (SMCFCS). The simplest possible method to include an interaction in the imputation model is to calculate it passively after imputing the component of the interaction separately; this is known as passive imputation. Von Hippel \textit{et al} (2009) showed that passive imputation will, in general, lead to bias and recommended another approach called ‘transform then impute’ (i.e., JAV), which considers the interaction as an independent variable in the imputation model (\cite{VonHippel2009HowVariables}). Seaman \textit{et al} (2014) demonstrated that unless JAV is used in special cases (i.e., for linear regression and where missing data are assumed missing completely at random) it can also lead to biased inference (\cite{Seaman2012MultipleMethods}). Seaman \textit{et al} further showed that JAV will lead to biases in logistic regression models when imputing a partially observed variable that has a non-linear (e.g., quadratic) effect. However, while results are expected to be similar, it is not known how passive or JAV imputation approaches perform for logistic regression models containing an interaction with a partially observed variable (i.e., $XZ$) (\cite{Seaman2012MultipleMethods}). Von Hippel \textit{et al} also proposed another approach to impute $X$: 'stratify, then impute' (we refer to this as Stratify-Impute-Append [SIA]). SIA imputes $X$ within strata (i.e., $Z=1$ and $Z=0$) of the fully observed variable in the interaction. Using the example before, this is to impute the partially observed cancer diagnostic stage variable within strata of the fully observed binary comorbidity covariate $Z$. \\
    
    More recently, Bartlett \textit{et al} (2014) developed an approach to perform multiple imputation by fully conditional specification (FCS) where the substantive model is a linear, logistic, or time-to-event model (\cite{Bartlett2014MultipleModel}). This approach accommodates complex terms in the substantive model (e.g., including interaction terms) whereby making the imputation model compatible with the substantive model: thus termed substantive model compatible fully conditional specification (SMCFCS). \\
    
    Since logistic regression models are a commonly used tool and often include interaction terms, it is important to identify the best way to handle missing data in these models appropriately. However, current research has not yet properly compared various imputation methods for this commonly occurring scenario. We aim to fill this gap, motivated by an applied example. Cancer stage at diagnosis, amongst other key variables such as cancer treatment, is a key prognostic indicator of survival. Similarly, for most cancers, other outcomes of interest are often investigated (e.g., the odds of death within a short term after diagnosis). The presence of comorbidities is known to influence the chance of a patient experiencing a delayed cancer diagnosis (i.e., a later stage at diagnosis). In epidemiological studies this effect is known as effect modification (i.e., interaction). Effect modification can be accounted for in an analysis by including an interaction term between the two variables. However, diagnostic stage is not always properly recorded and the probability of a missing record varies depending on the cancer type.\\
    
    In this article, we aim to identify and recommend a best-performing multiple imputation approach for various scenarios when the substantive model is a logistic regression model that has an interaction term containing a partially observed variable. In the Methods section we formally describe the imputation approaches and specifications of the simulations by following the ADEMP reporting scheme (\cite{Morris2019UsingMethods}). We also introduce an empirical example using England-based cancer registry data of patients diagnosed with diffuse large B-cell lymphoma, and we illustrate the application of each imputation method for this real-world setting. In the Results section, we show the performance (bias, coverage and relative error) of the imputation methods and the results of the empirical example. We end with a discussion on the performance and applicability of each imputation method.

\section{Methods}\label{Methods}

\paragraph{Introduction to the problem}
    \, \\
        Suppose data were observed on a sample of individuals for a binary outcome ($Y$), an exposure ($X$), and an additional covariate ($Z$). Suppose also that the data in $X$ was not fully observed (i.e., there is some missing data for some individuals). To ascertain the association between $Y$ and $X$, adjusted for $Z$, one could perform a logistic regression analysis on only those individuals who have a complete (fully observed) set of $X$ (i.e., complete case analysis). Assuming the missingness mechanism did not involve the outcome, e.g. if the data were \textit{missing completely at random}, these results would not be biased but would be potentially inefficient. If the data were \textit{missing at random} given the outcome, instead, the complete case analysis would be biased, and inefficient, compared to results after performing multiple imputation. \\
        
        Multiple imputation is often a useful technique to reduce bias and restore some efficiency when data are missing. However, the imputation model must be correctly specified and reflect the relationships between the variables in the substantive model. If the substantive model included an interaction between the partially observed variable ($X$) and the fully observed variable ($Z$), then this interaction must also be accounted for in the imputation model. We aimed to find the multiple imputation approach with the least bias, optimal coverage, and smallest relative error for handling missing data when the substantive analysis model is a logistic regression model that includes an interaction containing a partially observed variable. \\
    
    \paragraph{Data-generating mechanisms}
    \, \\
        Let $Y_{i}$, $X_{i}$ and $\mathbf{Z}_{i}$ denote the values of the binary random variables for the outcome, exposure, and covariates, respectively, for individual $i$ ($i = 1, ..., n$). Assume that $(Y_{1}, X_{1}, \mathbf{Z}_{1}), ..., (Y_{n}, X_{n}, \mathbf{Z}_{n})$ are independently and identically distributed. Let $R_{i} = 1$ if $X_{i}$ is observed (i.e. if subject $i$ is a complete case), with $R_{i} = 0$ otherwise. Let $n_{1}$ denote the number of complete cases. \\
        
        Let $Z_{1}$ represent treatment (1 = treated, 0 = untreated), $Z_{2}$ represents age (centred on the mean and rescaled by a factor of 10), $Z_{3}$ represents sex (1 = female, 0 = male), $Z_{4}$ is a unit increase in deprivation level (higher is more deprived), $Z_{5}$ represents comorbidity (1 = comorbidity, 0 = no comorbidity), and $X_{1}$ represents the cancer diagnostic stage (1 = late stage, 0 = early stage). These variables were simulated as $Z_{1} \sim Binomial(0.3)$, $Z_{2} \sim N(70, 10)$, $Z_{3} \sim Binomial(0.5)$, $Z_{4} \sim Beta(1,1.2)$, $Z_{5} \sim Binomial(0.3)$, and $X_{1} \sim Binomial(0.4)$. The outcome $Y$ was specified to mimick values estimated from real data and generated according to the following logistic regression model:
        
        \begin{equation}
            logit(\pi_{i}) = -3 + 0.85Z_{1i} + 1.3Z_{2i} + 0.9Z_{3i} + 1.2Z_{4i} + 0.9Z_{5i} + 1.4X_{1i} + 1.3Z_{5i}X_{1i} 
        \end{equation}    
        \\
        
        \noindent where $-3$ is the intercept ($\beta_{0}$). \\
        
        Missingness was imposed on $X_{1}$ using a missing at random (MAR) missing data mechanism. The probability of observing $X_{1}$ (i.e., $R_{i} = 1$) was dependent on the variables in the substantive model (including the outcome) with parameters defined as:
        
            $$ logit(\pi_{i}) = \alpha_{0} + \alpha_{1}Z_{1i} + \alpha_{2}Z_{2i} + \alpha_{3}Z_{3i} + \alpha_{4}Z_{4i} + \alpha_{5}Z_{5i} + \alpha_{6}Y_{i} $$ \,
        
        \noindent where each $\alpha_{k}$ ($k = 1, ..., 6$) was set to $1$. The intercept ($\alpha_{0}$) was set to make the probability of observing $X_{1}$ equal to 0.8 (i.e., 20\% missing data in diagnostic stage).  \\
    
        We define six data-generating mechanisms (DGM) where a factor is varied one-by-one away from a “base-case” data-generating mechanism (Table \ref{table:Table1}). The base-case DGM assumes the 20\% missing data in $X$ is \textit{missing at random}, the fully observed binary variable ($Z$) has a prevalence of 30\%, and their interaction (i.e., $XZ$) gives an odds ratio of 1.3 (i.e., amongst those with $Z=1$, the odds of $Y=1$ is 1.3 times higher amongst those with $X=1$ compared to those with $X=0$). Each DGM uses parametric draws from a known model where the true data-generating model is known. \\
        
        \begin{table}[ht]
        \centering
        \begin{threeparttable}
        \caption{Specifications of each data-generating mechanism.}
        \label{table:Table1}
        \begin{tabular}{lrlccccc}
            \hline
            & \textit{} &  & \multicolumn{5}{c}{\textbf{Specifications}} \\ \cline{4-8} 
            & \textit{} &  & \textbf{\begin{tabular}[c]{@{}c@{}}Missing data \\ mechanism\end{tabular}} & \textbf{\begin{tabular}[c]{@{}c@{}}Fully observed \\ variable (Z)\end{tabular}} & \textbf{\begin{tabular}[c]{@{}c@{}}Prevalence of \\ fully observed \\ variable (\%Z)\end{tabular}} & \textbf{\begin{tabular}[c]{@{}c@{}}Proportion of \\ missingness \\ (\%X)\end{tabular}} & \textbf{\begin{tabular}[c]{@{}c@{}}Effect of \\ interaction \\ (Odds ratio)\end{tabular}} \\ \cline{4-8}
            \multicolumn{2}{r}{\textbf{DGM\tnote{1}}} &  &  &  &  &  &  \\
             & \textit{(Base case) 1} &  & MAR\tnote{2} & Binary & 30\% & 20\% & 1.3 \\
             & \textit{2} &  & - & - & - & - & 1.1 \\
             & \textit{3} &  & - & - & - & - & 1.7 \\
             & \textit{4} &  & MCAR\tnote{2} & - & - & - & - \\
             & \textit{5} &  & - & Continuous & - & - & - \\
             & \textit{6} &  & - & - & 1\% & - & - \\ \hline
        \end{tabular}
            \begin{tablenotes}
                \item [1] Each data-generating mechanism (DGM) is the same as the 'base case' except for the specification in which there is a specified difference.
                \item [2] MAR: missing at random, MCAR: missing completely at random.
            \end{tablenotes}
        \end{threeparttable}
        \end{table}
        
        We simulated 10,000 observations ($n_{obs}$) containing data based on distributions found within real-world settings of cancer registry data on patients with diffuse large B-cell lymphoma in England (\cite{Smith2021AssociationStudy,Smith2021ExcessEngland,Smith2021InvestigatingEngland}). We chose a large enough sample of repetitions ($n_{sims} = 1000$) such that we obtained a small enough Monte Carlo standard error without unfeasible computational time. The formula for the Monte-Carlo confidence interval around the mean estimate is:\cite{Tang2005ATrial} \,
        
        $$ p \pm 1.96 * \sqrt{p(1-p)/B} $$  \,
        
        \noindent If we substitute $p$ with the nominal coverage probability (i.e., 0.95) and $B$ with the number of simulations ($ n_{sim} = 1,000$), the estimated coverage (see Section \ref{perfmeas}) should fall between 93.6\% and 96.4\% (i.e., approximately 19 times out of 20). \\
        
        The estimands are the exponential of the fixed effect parameter estimates (i.e., odds ratios) of the substantive logistic regression model (Equation 1) for (i) the fully observed variable ($Z_{5}$), (ii) the partially observed variable ($X_{1}$), and (iii) the interaction term ($Z_{5}X_{1}$). \\
        
    \paragraph{Imputation approaches (Methods)} \, \\
        \\
        For each simulated data set, and for each data-generating mechanism (DGM), we performed multiple imputation in the following manner: passive imputation, JAV, SIA, and finally SMC-FCS. We first briefly describe each approach. \\
        
        In passive imputation the imputation model is specified by the distribution of $X$ given $Y$ and $Z$. Missing values in $X$ are imputed from this distribution and the values for the interaction $XZ$ are then calculated. Passive imputation ensures that the imputed values of $XZ$ conform to the relationship between $X$ and $Z$ (i.e., that $XZ = X \times Z$. On the other hand, JAV imputation does not ensure this relationship holds. JAV imputation ignores the relationship $X$ and $Z$ and imputes $XZ$ as just another variable. This approach imputes missing values assuming that $Y, X, Z$, and $XZ$ are separate variables. Therefore, imputed values of $XZ$ will not always be consistent with $X \times Z$ (i.e., for an individual with $Z=1$, $X$ might be imputed as 1, giving $XZ = 1$, but $XZ$ might be imputed as 0). \\
        
        As an alternative to handle the interaction term, one could use Stratify-Impute-Append (SIA) This approach imputes values of $X$ within levels of the $Z$. Imputing separately in levels of $Z$ allows associations between $X$ and $Y$ to differ according to the level of $Z$. The three approaches mentioned thus far either ignore the compatibility between the imputation model and the substantive model (i.e., passive and JAV) or aim to by-pass the interaction term (i.e., SIA). The last approach, substantive model compatible full conditional specification (SMCFCS) ensures $X$ is imputed using an imputation model that is compatible with the substantive model. The compatibility is achievable through rejection sampling, which involves drawing values from a proposal density (i.e., a function of the missing values in $X$ given the other variables) and accepting the draw if it satisfies certain conditions based on a ratio of a target density to the proposal density. The target density incorporates the parameters from the substantive model, allowing for compatibility with the imputation model. \\
        
        For DGM 1-4, and DGM 6 (i.e., not including DGM 5), firstly, missing values of $X$ were passively imputed and values for the interaction ($XZ$) were calculated from the combination of the imputed $X$ values and the observed $Z$ values. Secondly, we imputed missing $XY$ values using the 'just another variable' (JAV) approach, as proposed by von Hippel (\cite{VonHippel2009HowVariables}). Thirdly, we performed Stratify-Impute-Append (SIA) imputation by initially stratifying the data by the groups of the fully observed variable ($Z$) within the interaction ($XZ$). We then imputed missing values in $X$, separately, for each stratified data set, and calculated the values for interaction ($XZ$). The stratified data sets were then appended for calculation of the estimands. Fourthly, we used SMC-FCS to impute missing values in $X$ and $XZ$.   \\
        
        For DGM 5, the fully observed variable ($Z$) in the interaction ($XZ$) was of a continuous form, and passive imputation, JAV, and SMC-FCS were performed as in DGM 1-3 (and DGM 6). However, for SIA, the continuous form of $Z$ was categorised using quintiles (i.e., creating 5 groups of equal sizes). SIA imputation for $X$ was carried out in 5 separate data sets (i.e., one data set for each of the five groups), which were then combined. The odds ratios (i.e., estimands) were calculated assuming a linear distribution (i.e., continuous form) across the levels of the categorical variable $Z$. \\

        As a comparison, we also include a "null scenario" where the substantive model does not include an interaction term (i.e., where the effect of the interaction has an odds ratio of 1).\\
        
        For each imputation method we impute 10 data sets with 10 iterations between imputations. Estimates and confidence intervals (CI) were calculated using Rubin's rules (\cite{Little1987StatisticalData,Rubin1987MultipleSurveys}). We used R software for all analyses. The Passive, JAV, and SIA imputation methods were performed using the \textit{mice} package (\cite{vanBuuren2011Mice:R}) and SMCFCS was performed using the \textit{smcfcs} package (\cite{Bartlett2014MultipleModel}). \\
        
    \paragraph{Performance measures}\label{perfmeas} \, \\
    \\
        We assess the performance of each of the four imputation methods by comparing: \\
        
        \noindent (i) relative bias (i.e., the relative difference between the mean estimated coefficient, $E[\hat{\theta}],$ and the true value of the coefficient, $\theta$) \\
        
        $$ Relative Bias \quad = \quad \frac{E[\hat{\theta}]-\theta}{\theta} \quad = \quad \frac{\frac{1}{n_{sim}}\sum_{i=1}^{n_{sim}} \hat{\theta}_{i}-\theta}{\theta} $$ \,
        
        \noindent (ii) coverage (i.e., proportion of 95\% confidence intervals that include the true coefficient) \\
        
        $$ Coverage \quad = \quad {Pr}\left(\hat{\theta}_{low} \leq \theta \leq \hat{\theta}_{upp}\right) \quad  =  \quad \frac{1}{n_{sim}} \sum_{i=1}^{n_{sim}} 1\left(\hat{\theta}_{low, i} \leq \theta \leq \hat{\theta}_{upp,i}\right) $$ \,
        
        \noindent (iii) relative error is a comparison between the average model-based standard error (ModSE) and the empirical standard error (EmpSE) (\cite{Burton2006TheStatistics}), and expresses how closely the model-based standard error approximates the empirical standard error (EmpSE):\\
        
        $$ \widehat{ModSE} \quad = \quad \sqrt{\frac{1}{n_{sim}} \sum_{i=1}^{n_{sim}} \widehat{Var}\left(\hat{\theta}_{i}\right)} $$ \\
        
        $$ \widehat{\mathrm{EmpSE}} \quad = \quad \sqrt{\frac{1}{n_{sim}-1} \sum_{i=1}^{n_{sim}}\left(\hat{\theta}_{i}-\bar{\theta}\right)^{2}} $$ \\
        
        $$ Relative error \quad = \quad 100\left(\frac{\widehat{ModSE}}{\widehat{EmpSE}}-1\right) $$ \\

    \subsection{Analysis of data from Cancer Registry records}\label{Analysis}

    The National Cancer Registry and Analysis Service (NCRAS), run by Public Health England (PHE), is responsible for cancer registration in England and supports cancer epidemiology, public health, service monitoring and research. Information on patients diagnosed with cancer is essential for assessing the public health system's ability to care for these patients. In England, 17,345 patients were diagnosed with diffuse large B-cell lymphoma between 1st January 2014 and 31st December 2017. Information was available on the patient's age at diagnosis, sex, socioeconomic status, comorbidity status, and cancer stage at diagnosis. \\
    
    A patient's comorbidity status is based on the Charlson comorbidity index (\cite{Charlson1987AValidation}) and was defined as “the existence of disorders, in addition to a primary disease of interest, which are causally unrelated to the primary disease” (\cite{Porta2014AEpidemiology,Feinstein1970TheDisease}). Comorbidities were coded within Hospital Episode Statistics (\cite{NHSDigital2015HospitalStatistics}) data according to the International Classification of Diseases, 10th revision (\cite{InternationalAgencyforResearchonCancer2013InternationalOncology}). Patients with any previous malignancy were removed. For each patient, we defined a time window of 6–24 months prior to cancer diagnosis for a comorbidity to be recorded. A patient’s comorbidity status was determined using an algorithm developed by Maringe et al (\cite{Maringe2017ReproducibilityComorbidities}). \\
        
    It is known that the presence of comorbidity symptoms can delay or hasten the cancer diagnostic stage (\cite{Renzi2019ComorbidMechanisms}). For example, comorbidity symptoms similar to the cancer might hide the cancer symptoms, whereas dissimilar symptoms might highlight the differing diagnoses (particularly if the patient is frequently attending routine health appointments for their underlying comorbid condition). For patients with DLBCL, it has been shown that comorbidities are associated with diagnostic delay (\cite{Smith2021AssociationStudy,Smith2021InvestigatingEngland,Smith2022MediatingEngland}). It would be natural to assume that a statistical model included an interaction between comorbidity and cancer diagnostic stage when investigating the probability of short-term mortality. \\
    
    We aimed to estimate the parameters of a logistic regression model for the odds of death within 90 days since cancer diagnosis amongst patients diagnosed with diffuse large B-cell lymphoma. Data was available on patient's age (continuous, range: 15.1 - 99.0), sex (binary, Male = 0, Female = 1), deprivation (continuous, range: -1.9 - 6.7), comorbidity (None = 0, at least one = 1), and diagnostic stage (early [stages I/II] = 0, late [stages III/IV] = 1). Information on treatment was not available. Patient characteristics, and the crude odds ratio (OR) of death within 90 days since diagnosis is shown in table \ref{tab:Table2}. \\
    
    For the adjusted analysis, the substantive model adjusted for age, sex, deprivation, comorbidity, diagnostic stage, and an interaction between comorbidity and diagnostic stage. Multiple imputation was performed using the four imputation methods (i.e., passive imputation, JAV, SIA, and SMCFCS). We used 10 imputation with 10 iterations and combined results using Rubin's rules. We tabulate the results of each imputation method in the results section along with the results from a complete case analysis (i.e., ignoring missing data). \\
    
    \begin{table}[ht]
    \centering
    \begin{threeparttable}
    \caption{Patient characteristics and crude odds ratios of death within 90 days after diagnosis of diffue large B-cell lymphoma amongst 17,345 patients diagnosed in England between 2014 and 2017.\\} 
    \label{tab:Table2}
    \begin{tabular}{lrrrrrr}
    \textbf{} &
      \textit{} &
      \multicolumn{1}{c}{\textbf{\begin{tabular}[c]{@{}c@{}}Alive\\ N (\%)\end{tabular}}} &
      \multicolumn{1}{c}{\textbf{\begin{tabular}[c]{@{}c@{}}Dead\\ N (\%)\end{tabular}}} &
      \multicolumn{1}{c}{\textbf{Crude OR}} &
      \multicolumn{1}{c}{\textbf{95\% CI}} &
      \multicolumn{1}{c}{\textbf{p-value}} \\ \cline{3-7} 
    \textbf{Age\tnote{1}}      & \textit{}                 &                        &                       &              &              &                  \\
    \textbf{}         & \textit{years (SD)}       & 67.4 (14.5)            & 76.6 (11.3)           & 1.83         & 1.78 - 1.91  & \textless{}0.001 \\
    \textbf{Sex}      & \textit{}                 &                        &                       &              &              &                  \\
    \textbf{}         & \textit{Male}             & 8,025 (55.5)           & 1,578 (54.8)          & Ref          &              &                  \\
    \textbf{}         & \textit{Female}           & 6,439 (44.5)           & 1,303 (45.2)          & 1.03         & 0.95 - 1.12  & 0.484            \\
    \textbf{Deprivation}      & \textit{}                 &                        &                       &              &              &                  \\
    \textbf{}         & \textit{years (SD)}       & -0.02 (1.44)           & 0.12 (1.50)           & 1.07         & 1.04 - 1.10  & \textless{}0.001 \\
    \multicolumn{2}{l}{\textbf{Comorbidity}}      &                        &                       &              &              &                  \\
    \textbf{}         & \textit{None}             & 6,009 (41.5)           & 808 (28.1)            & Ref          &              &                  \\
    \textbf{}         & \textit{At least one}     & 8,455 (58.5)           & 2,073 (72.0)          & 1.82         & 1.67 - 1.99  & \textless{}0.001 \\
    \multicolumn{2}{l}{\textbf{Diagnostic stage}} &                        &                       &              &              &                  \\
    \textbf{}         & \textit{Early}            & 5,193 (39.4)           & 509 (22.5)            & Ref          &              &                  \\
    \textbf{}         & \textit{Late}             & 7,986 (60.6)           & 1,751 (77.5)          & 2.24         & 2.01 - 2.48  & \textless{}0.001 \\
    \textbf{}         & \textit{Missing\tnote{2}}          & \textit{1,285 (8.9\%)} & \textit{621 (21.6\%)} & \textit{n/a} & \textit{n/a} & \textit{n/a}     \\ \hline
    \end{tabular}
        \begin{tablenotes}
        \item SD: standard deviation, OR: odds ratio, 95\% CI: confidence interval
        \item [1] Age is centred on its mean (68.9 years) and rescaled by a factor of 10
        \item [2] Missing proportions are calculated separately from observed data.
        \end{tablenotes}
    \end{threeparttable}
    \end{table}
    
    \newpage \, \\
    \newpage

    \section{Results}\label{Results}

    \subsection{Simulation results}
    
        The results of the performance measures for each imputation approach when imputing the interaction term are shown in the Appendix table \ref{tab:TableA1}. We report the results of each of the DGMs by the relative bias, coverage, and relative error in the following sections.\\
        
        \noindent In the null scenario (i.e., no interaction), there was negligible bias, and optimal coverage, of the fully and partially observed variables for all imputation methods (Appendix table \ref{tab:TableA1}). The relative error was similar across all four imputation methods for the fully observed variable but SMCFCS had the lowest relative error for the partially observed variable. \\

    \subsubsection{Bias (and relative bias)}
    
\begin{figure}[ht]
            \centering
            \includegraphics[scale=0.55]{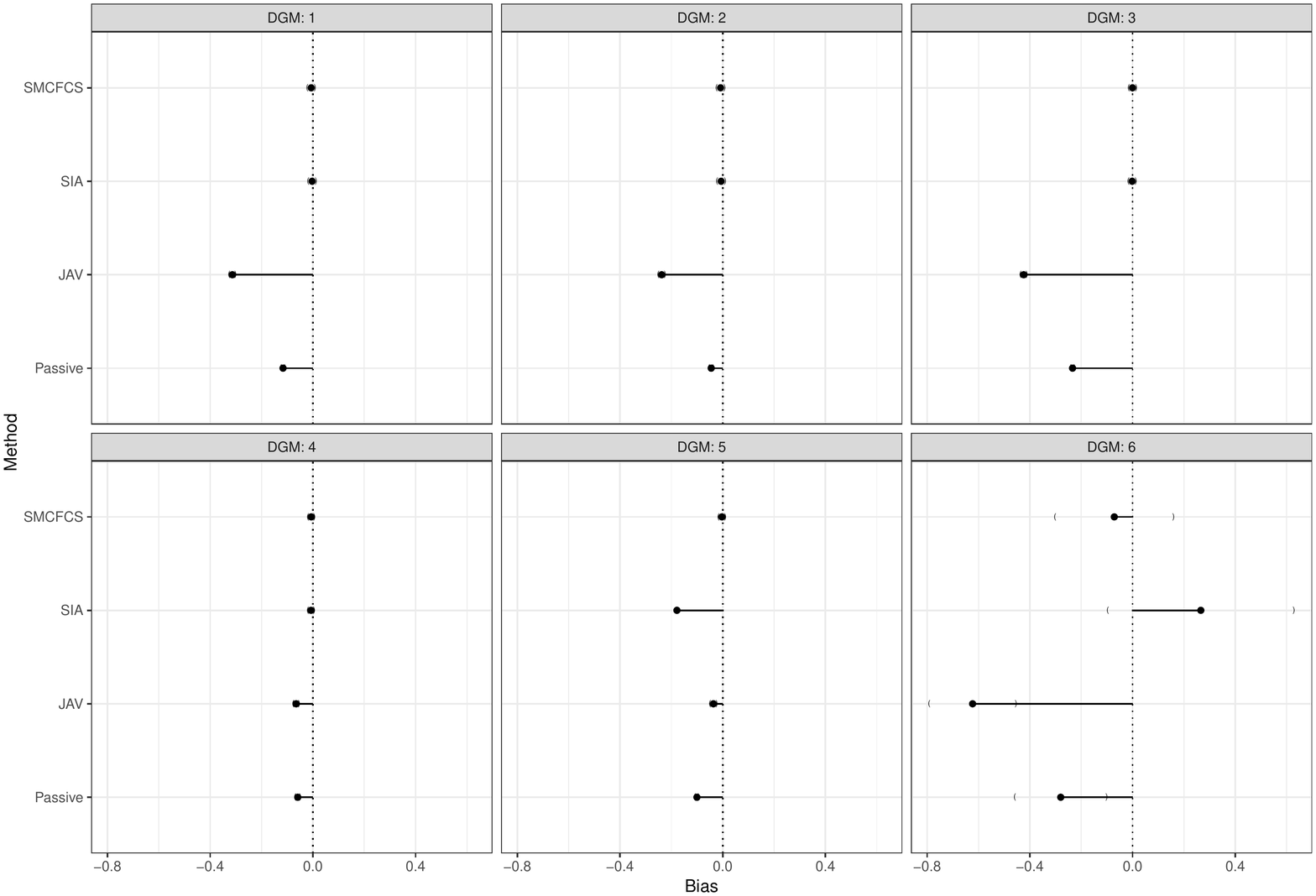}
            \caption{Bias of each imputation method for the interaction term across simulations (n=1,000) by data-generating mechanism.}
            \label{fig:1}
        \end{figure}

        Figure \ref{fig:1} shows bias for the coefficient of the interaction for each imputation method, The Appendix table \ref{tab:TableA1} shows the precise values. For DGMs 1-4, SIA and SMCFCS showed negligible bias, JAV was severely and the most biased, and passive imputation was biased for DGMs 1-4. In DGM 5 (continuous form of fully observed variable), SMCFCS showed negligible bias, SIA was the least accurate (most biased) followed by passive imputation, then JAV. In DGM 6, JAV had the largest bias, followed by passive imputation and SIA, and SMCFCS was the least biased. \\
        
\begin{figure}[ht]
            \centering
            \includegraphics[scale=0.55]{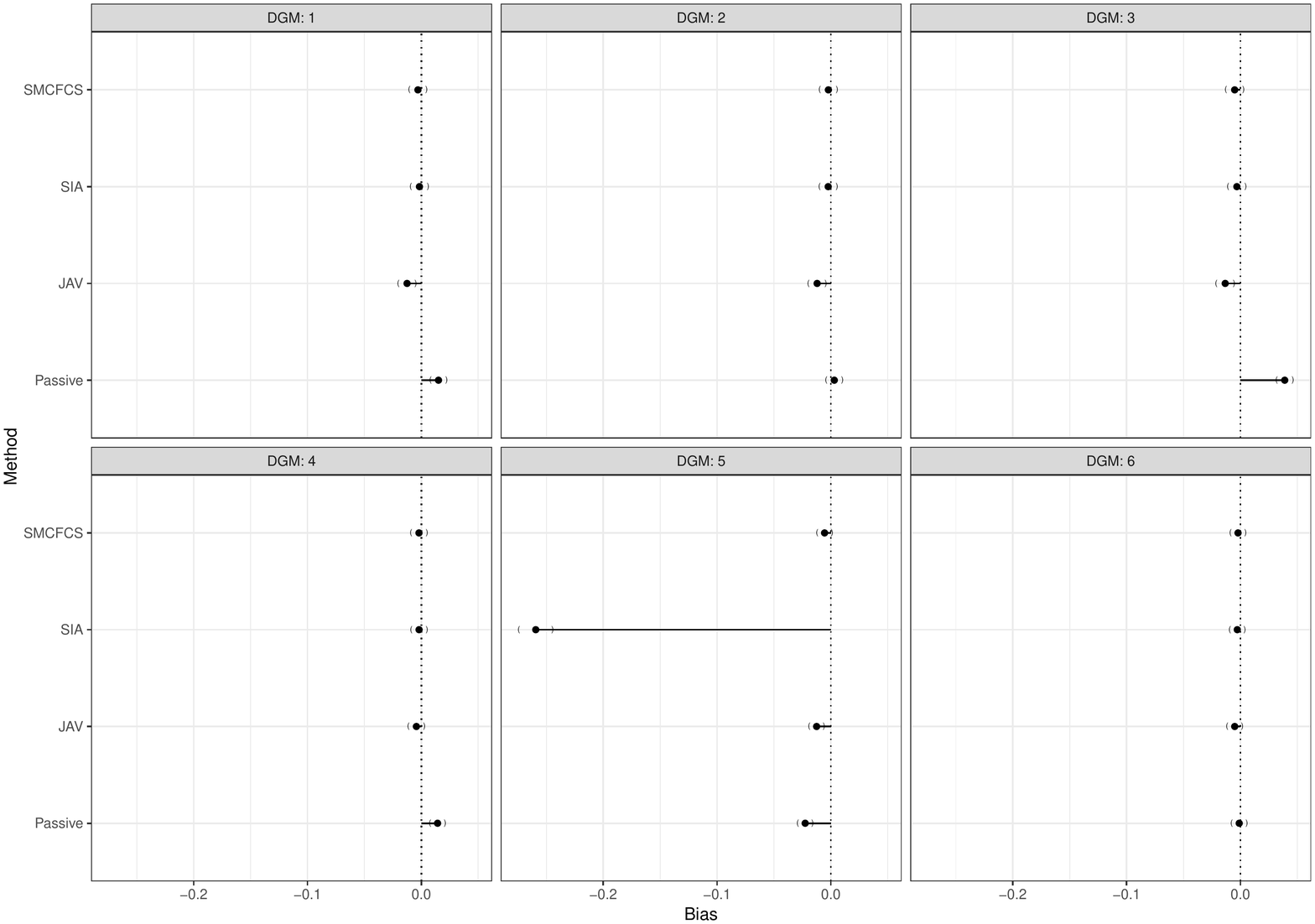}
            \caption{Bias of each imputation method for the partially observed variable across simulations (n=1,000) by data-generating mechanism.}
            \label{fig:2}
        \end{figure}

        Figure \ref{fig:2} shows bias for the coefficient of the partially observed variable for each imputation method. For all four imputation approaches, there was negligible bias, except for SIA, which severely underestimated the effect when the partially observed variable had an underlying continuous form (i.e., DGM 5). Passive imputation, was slightly biased in DGMs 1, 3, 4, and 5. JAV showed an underestimate of the partially observed variable in DGMs 1, 2, 3, and 4. \\
        
\begin{figure}[ht]
            \centering
            \includegraphics[scale=0.55]{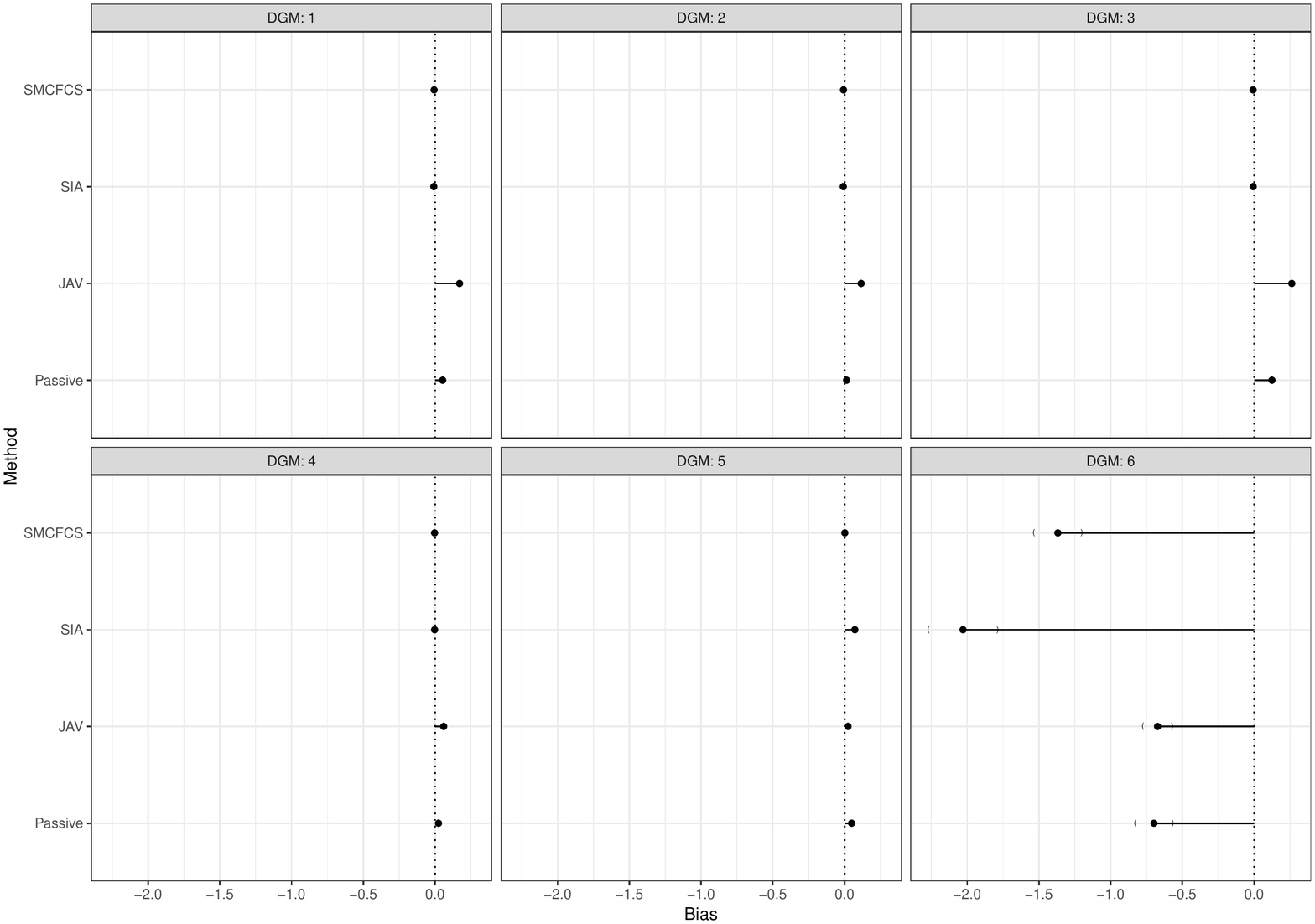}
            \caption{Bias of each imputation method for the fully observed variable across simulations (n=1,000) by data-generating mechanism.}
            \label{fig:3}
        \end{figure}

        Figure \ref{fig:3} shows the results of the bias for the coefficient of the fully observed variable for each imputation method. SIA and SMCFCS showed negligible bias for DGMs 1-5, but were the most biased for DGM 6. JAV consistently overestimated the effect of the fully observed variable in DGMs 1-4 but severely underestimated the effect in DGM 6 (low prevalence of the fully observed variable).\\
        
        \newpage \, 
        \newpage \, 
        \newpage \, 
        \newpage

\subsubsection{Coverage} 
        
\begin{figure}[ht]
            \centering
            \includegraphics[scale=0.55]{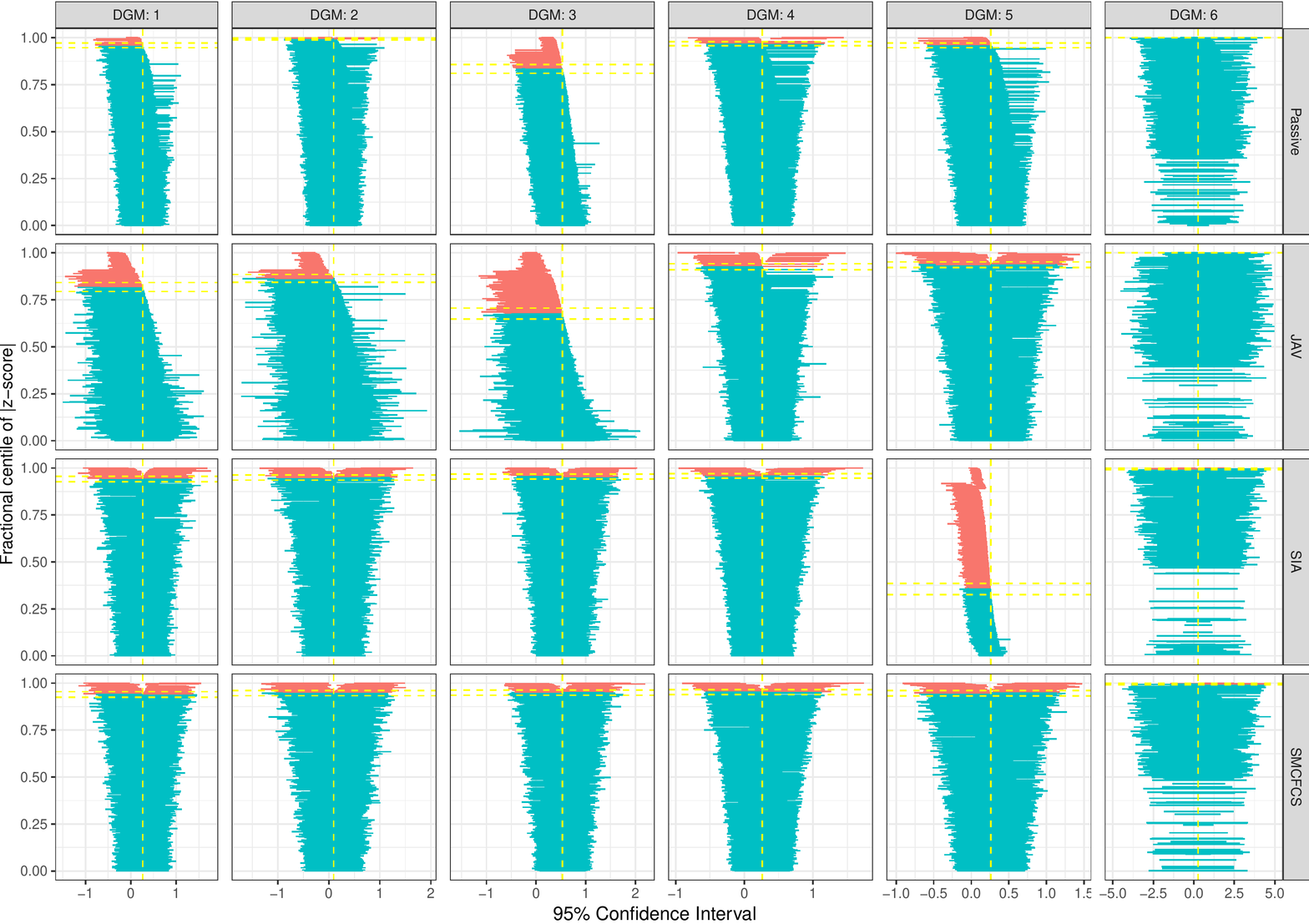}
            \caption{Interaction: Coverage of 95\% confidence intervals for the interaction term across simulations (n=1,000) by data-generating mechanism and for each imputation method.}
            \label{fig:4}
        \end{figure}

        Figure \ref{fig:4} shows the results of the coverage of 95\% confidence intervals for the interaction variable for each imputation method. Specific values are shown in the Appendix table \ref{tab:TableA1}. SMCFCS consistently showed an optimal coverage for DGMs 1-5, and was only affected by overcoverage for DGM 6. SIA had similar results, but for DG5, where it had severe undercoverage (35.6\%). Coverage for the other two methods was instead inferior, with JAV undercovering in all scenarios but DGM 6 and passive imputation showing overcoverage for DGM 2 and 6, and undercoverage for DGM3. \\
        
\begin{figure}[ht]
            \centering
            \includegraphics[scale=0.55]{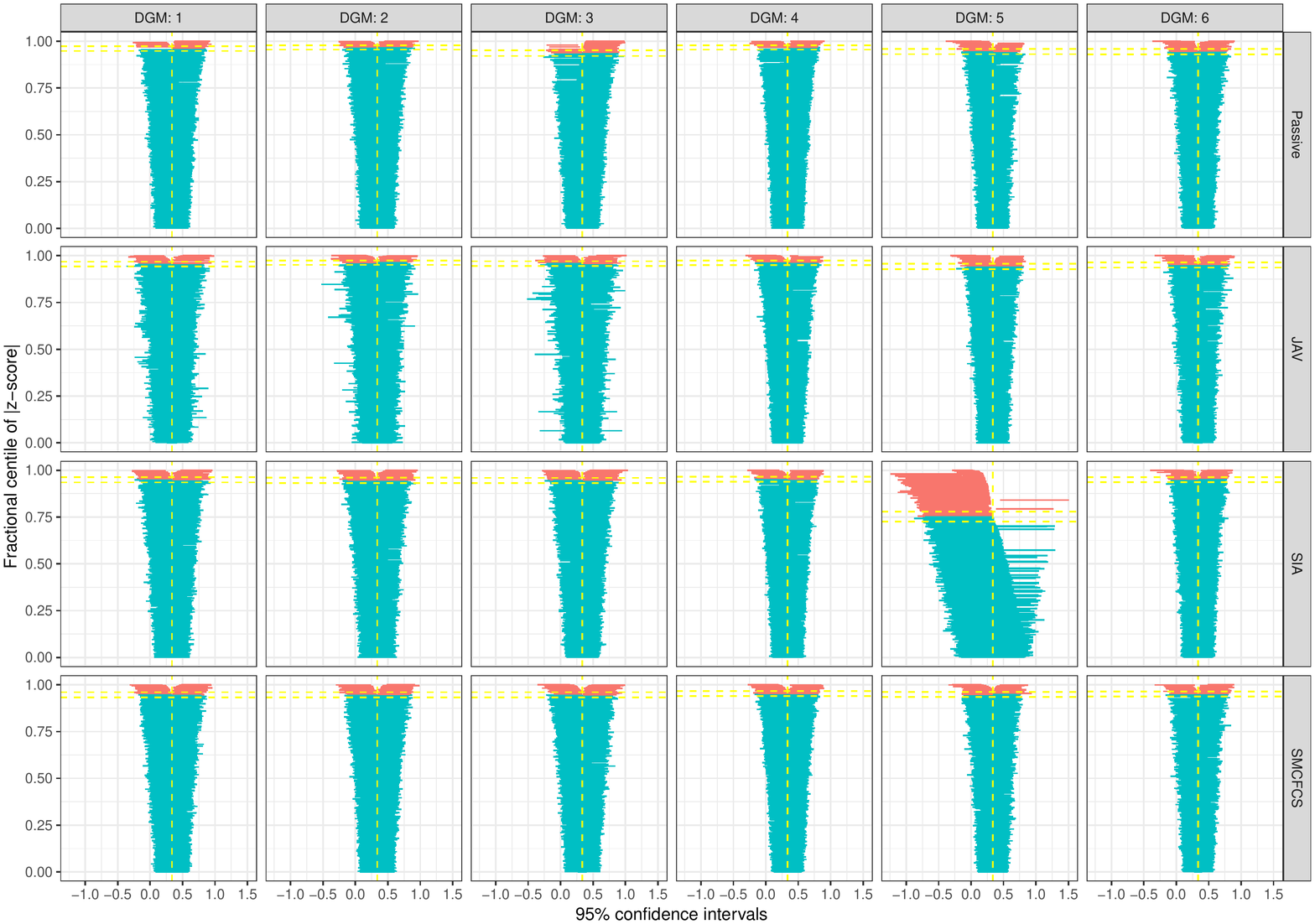}
            \caption{Stage: Coverage of 95\% confidence intervals for the partially observed variable across simulations (n=1,000) by data-generating mechanism and for each imputation method.}
            \label{fig:5}
        \end{figure}

        For the partially observed variable, all of the imputation approaches, except SIA in DGM 5, showed a similar coverage in the optimal range (93.6-96.4\%) (figure \ref{fig:5}). SIA had severe undercoverage (75.3\%) when the fully observed variable had an underlying continuous form (i.e., DGM5). \\
        
\begin{figure}[ht]
            \centering
            \includegraphics[scale=0.55]{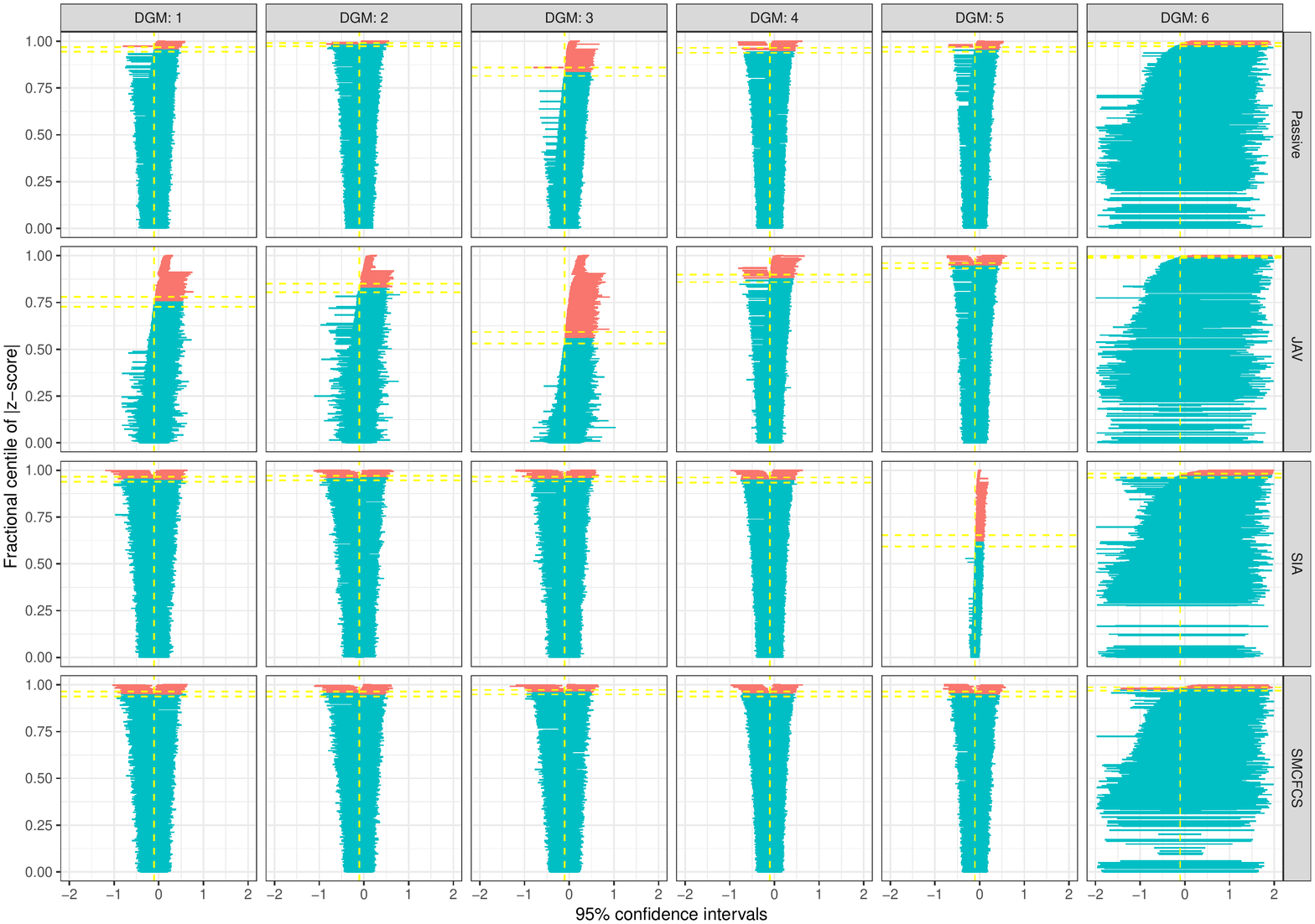}
            \caption{Comorbidity: Coverage of 95\% confidence intervals for the fully observed variable across simulations (n=1,000) by data-generating mechanism and for each imputation method.}
            \label{fig:6}
        \end{figure}

        For the fully observed variable (figure \ref{fig:6}), results were broadly in line with those for the interaction parameter. \\
        
        \newpage \,
        \newpage \,
        \newpage

    \subsubsection{Relative error} 
    
\begin{figure}[ht]
            \centering
            \includegraphics[scale=0.55]{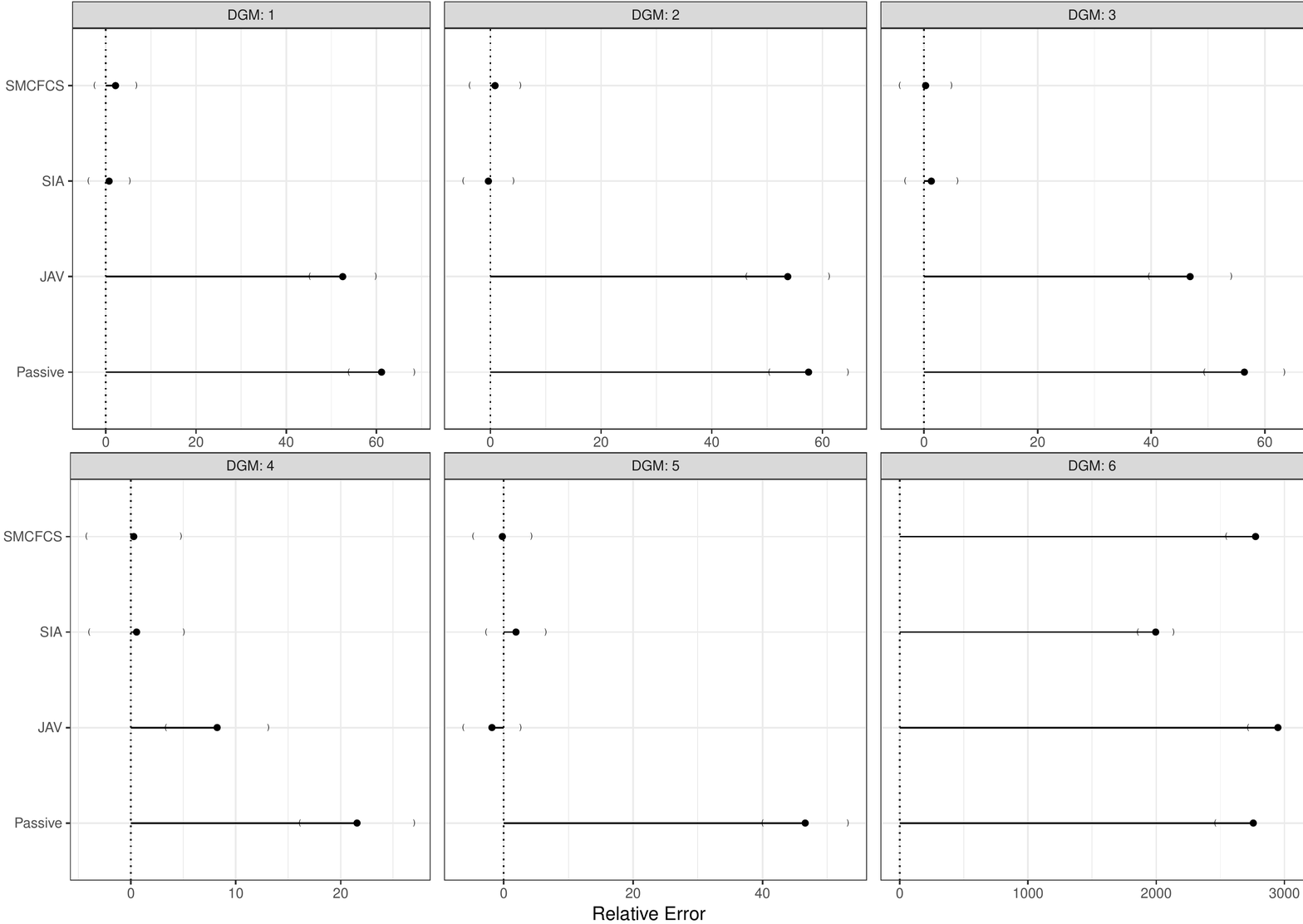}
            \caption{Relative error for the interaction term across simulations (n=1,000) by data-generating mechanism and for each imputation method.}
            \label{fig:7}
        \end{figure}

\begin{figure}[ht]
            \centering
            \includegraphics[scale=0.55]{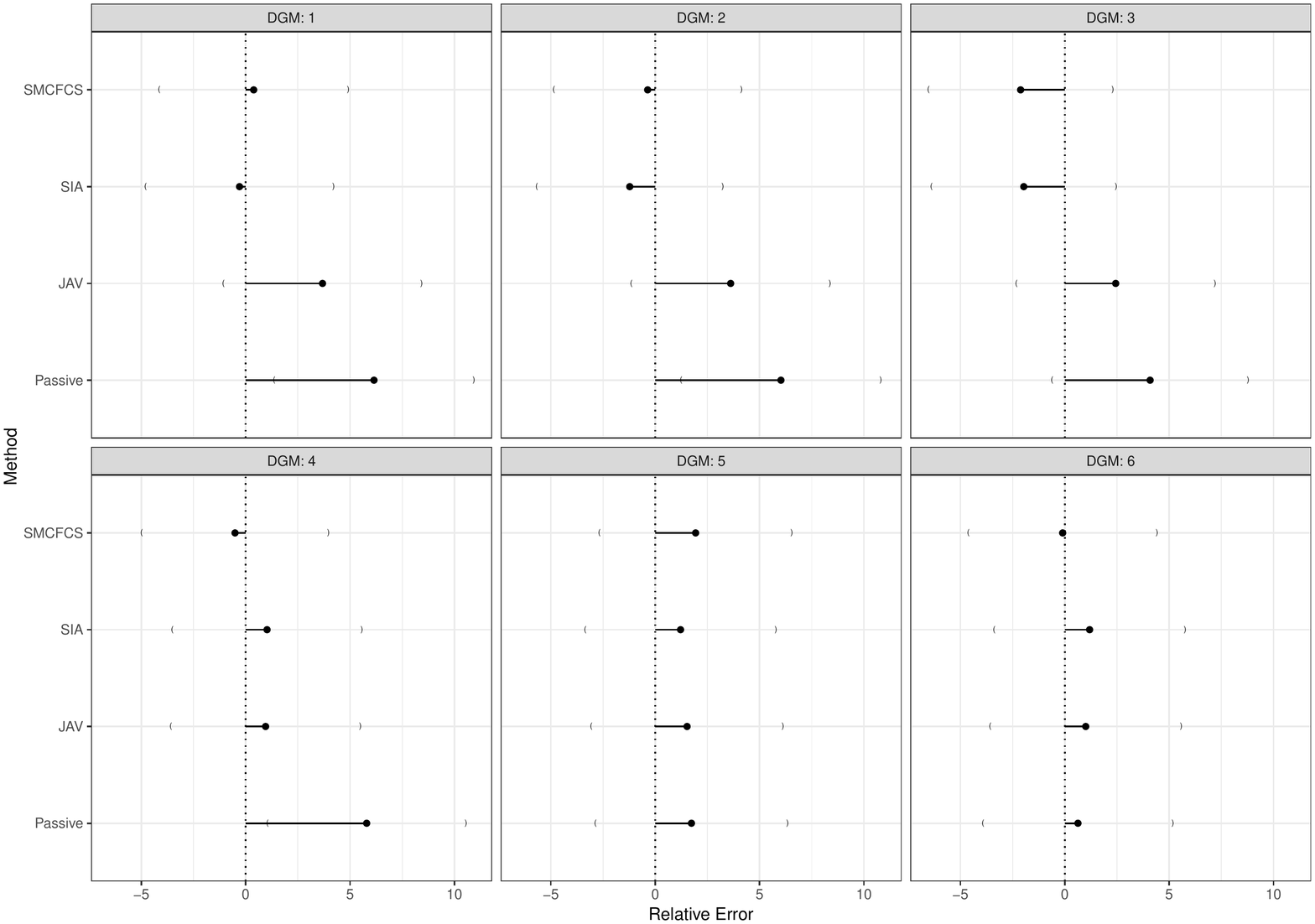}
            \caption{Relative error for the partially observed variable across simulations (n=1,000) by data-generating mechanism and for each imputation method.}
            \label{fig:8}
        \end{figure}

\begin{figure}[ht]
            \centering
            \includegraphics[scale=0.55]{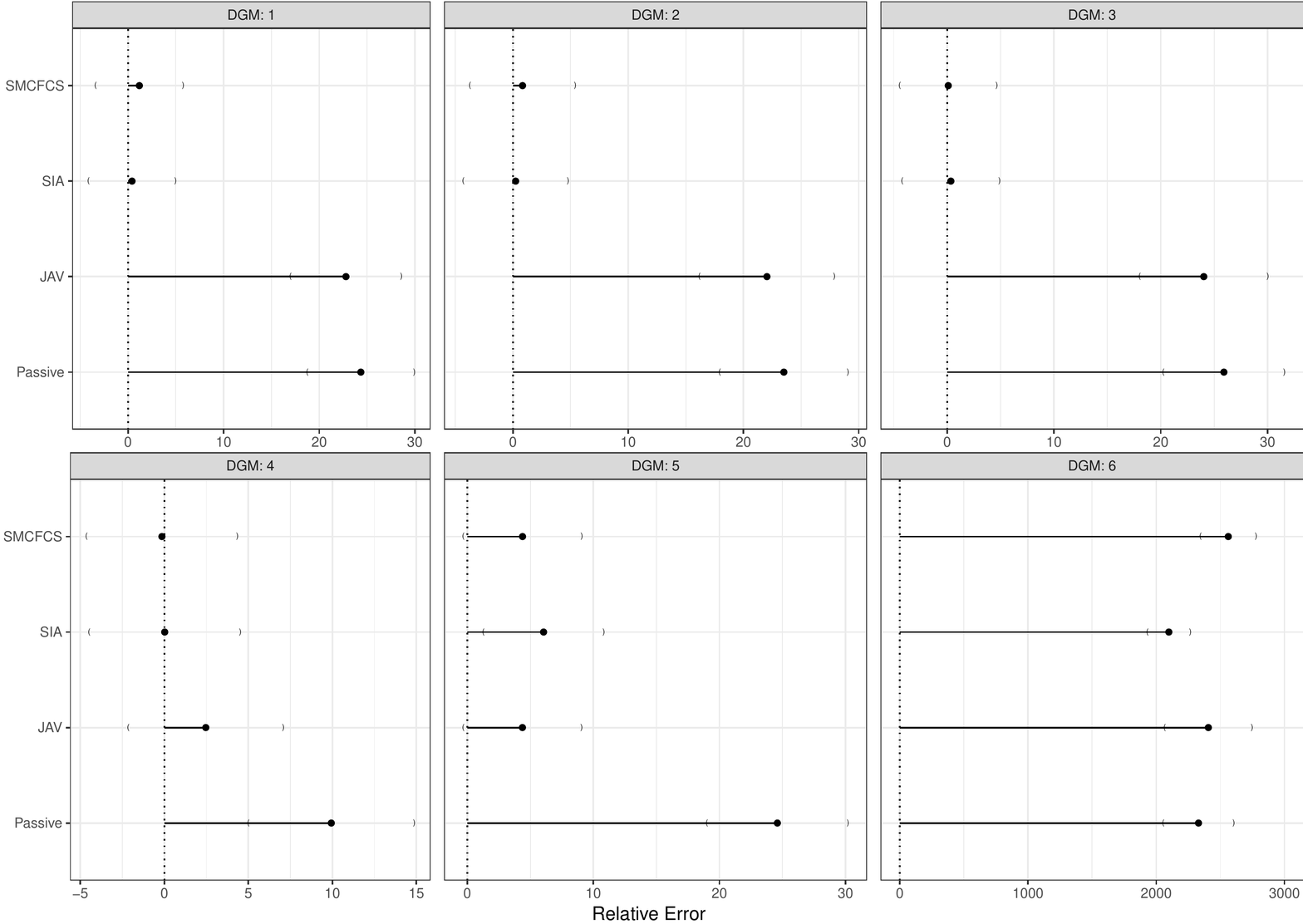}
            \caption{Relative error for the fully observed variable across simulations (n=1,000) by data-generating mechanism and for each imputation method.}
            \label{fig:9}
        \end{figure}

        The relative error for the interaction term, partially observed variable, and fully observed variable are shown in figures \ref{fig:7}, \ref{fig:8}, and \ref{fig:9}, respectively. \\
        
        For the interaction term (figure \ref{fig:7}), SMCFCS and SIA showed an optimal relative error (i.e., within the confidence intervals of the MCSE) for DGMs 1-5, but had a very large relative error for DGM 6 (low prevalence of the fully observed variable). JAV's average model-based standard error overestimated the empirical standard error (i.e., large relative error) within DGMs 1-3. JAV showed an optimal relative error for DGM 4 and DGM 5, and severely overestimted the relative error in DGM 6. Passive imputation showed the largest overestimate of the relative error for all DGMs. \\
        
        For the partially observed variable (figure \ref{fig:8}), SMCFCS, SIA and JAV showed a relative error that was within the expected range (i.e., within MCSE confidence intervals) for all DGMs. Passive imputation showed a slightly larger relative error than expected (i.e., outside the MCSE confidence intervals) for DGMs 1, 2, and 4, but was within MCSE confidence intervals for DGMS 3, 5, and 6.\\
        
        For the fully observed variable (figure \ref{fig:9}), SMCFCS and SIA showed negligible relative error for DGMs 1-4, a larger relative error for DGM 5, and a very large relative error for DGM 6. JAV showed a large relative error for DGMs 1-3, a slightly larger ModSE for DGMs 4 and 5, and a very large relative error for DGM 6. Passive imputation showed a large relative error for all DGMs.\\
        
        \newpage \,
        \newpage \,
        \newpage

    \subsection{Analysis of cancer registry data}
    
        Table \ref{table:realdata} shows the results of the analysis on the probability of death within 90 days since cancer diagnosis. Under complete case analysis, the interaction term (Odds Ratio [OR] 0.90, 95\% confidence interval [CI] 0.71 - 1.14) indicated a possible protective effect on the odds of death within 90 days amongst those with comorbidity and late stage at diagnosis but significant uncertainty remained. Amongst those with no comorbidities, the odds of death within 90 days was 2.51 (95\% CI: 2.04 - 3.07) times higher for those with a late stage at diagnosis compared to those with an early stage. Amongst those with early diagnostic stage, the presence of comorbidity increased the odds of death by 1.54 (95\% CI 1.25 - 1.90) compared to those without comorbidity. For the other variables in the model, the odds of death within 90 days was associated with age (OR 1.93, 95\% CI 1.84 - 2.02), sex (OR 0.90, 95\% 0.82 - 0.99), and deprivation (OR 1.12, 95\% CI 1.09 - 1.16), adjusting for the other covariates.\\
        
        The data from this example compares most closely with DGM 2 from the simulation study. The missing data in stage is assumed to be missing at random, the fully observed variable is of binary form, the prevalence of the fully observed variable is not small (i.e., 60.7\%), and there is a \textit{small} effect of the interaction between comorbidity and diagnostic stage on the probability of death. \\
        
        For the interaction term, SIA and SMCFCS showed very similar coefficients and confidence intervals, differing only by at most 2 decimal places. Passive imputation showed a slightly weaker effect of the interaction. JAV showed a strong protective effect (OR 0.73, 95\% CI 0.47 - 1.08) of the interaction term. The coefficient for the partially observed variable (diagnostic stage) was similar between SMCFCS and SIA, which differed from a similar result between JAV and passive imputation. The coefficient, and confidence intervals, for the fully observed variable (comorbidity status) was again similar between SIA and SMCFCS, but also similar to passive imputation. JAV showed a much larger harmful effect of comorbidity (OR 1.77, 95\% CI 1.28 - 2.44) compared to the other imputation approaches and the complete case analysis. \\
        
        There was negligible difference between the imputation approaches for the odds of death, and the confidence intervals (CI), for age, sex and deprivation and the substantial conclusions remained the same irrespective of the methods used. \\
        
        \begin{sidewaystable}
            \centering
            \begin{tabular}{rrrrrrrrrrr}
             &  &  & \multicolumn{8}{c}{\textbf{After multiple imputation}} \\ \cline{4-11} 
             & \multicolumn{2}{c}{\textbf{Complete Case Analysis}} & \multicolumn{2}{c}{\textbf{Passive}} & \multicolumn{2}{c}{\textbf{JAV}} & \multicolumn{2}{c}{\textbf{SIA}} & \multicolumn{2}{c}{\textbf{SMCFCS}} \\  \cline{2-11}
             & \textbf{OR} & \textbf{95\% CI} & \textbf{OR} & \textbf{95\% CI} & \textbf{OR} & \textbf{95\% CI} & \textbf{OR} & \textbf{95\% CI} & \textbf{OR} & \textbf{95\% CI} \\
            \multicolumn{1}{l}{\textbf{Age*}} & 1.93 & (1.84, 2.02) & 1.88 & (1.80, 1.96) & 1.87 & (1.79, 1.94) & 1.88 & (1.80, 1.96) & 1.88 & (1.80, 1.96) \\
            \multicolumn{1}{l}{\textbf{Sex}} &  &  &  &  &  &  &  &  &  &  \\
            \multicolumn{1}{r}{\textit{Male}} & Ref & - & Ref & - & Ref & - & Ref & - & Ref & - \\
            \multicolumn{1}{r}{\textit{Female}} & 0.90 & (0.82, 0.99) & 0.92 & (0.85, 1.00) & 0.92 & (0.85, 1.00) & 0.92 & (0.85, 1.00) & 0.92 & (0.85, 1.00) \\
            \multicolumn{1}{l}{\textbf{Deprivation}} & 1.12 & (1.09, 1.16) & 1.12 & (1.09, 1.15) & 1.12 & (1.09, 1.15) & 1.12 & (1.09, 1.15) & 1.12 & (1.09, 1.15) \\
            \multicolumn{1}{l}{\textbf{Comorbidity}} &  &  &  &  &  &  &  &  &  &  \\
            \multicolumn{1}{r}{\textit{None}} & Ref & - & Ref & - & Ref & - & Ref & - & Ref & - \\
            \multicolumn{1}{r}{\textit{At least one}} & 1.54 & (1.25, 1.90) & 1.43 & (1.17, 1.75) & 1.77 & (1.28, 2.44) & 1.45 & (1.18, 1.80) & 1.46 & (1.18, 1.80) \\
            \multicolumn{1}{l}{\textbf{Diagnostic stage}} &  &  &  &  &  &  &  &  &  &  \\
            \multicolumn{1}{r}{\textit{Early}} & Ref & - & Ref & - & Ref & - & Ref & - & Ref & - \\
            \multicolumn{1}{r}{\textit{Late}} & 2.51 & (2.04, 3.07) & 2.42 & (1.97, 2.97) & 2.43 & (1.85, 3.19) & 2.48 & (2.01, 3.07) & 2.49 & (2.03, 3.05) \\
            \multicolumn{1}{l}{\textbf{Interaction**}} & 0.90 & (0.71, 1.14) & 0.93 & (0.73, 1.18) & 0.72 & (0.47, 1.08) & 0.91 & (0.71, 1.17) & 0.91 & (0.71, 1.16)
            \end{tabular}
            \caption{Odds of mortality within 90 days from diagnosis of patients diagnosed with diffuse large B-cell lymphoma in England between 2014 and 2017. \textbf{JAV}: Just-Another Variable. \textbf{SIA}: Stratify-Impute-Append. \textbf{SMCFCS}: substantive model compatible fully conditional specification.\\
            *Increase in odds for each 10-year increase in age. **Interaction between stage and comorbidity.}
            \label{table:realdata}
        \end{sidewaystable}
    
    \newpage

\section{Discussion}\label{Discuss}

    We assessed the performance of four imputation approaches when handling missing data before fitting a logistic regression model with an interaction that contains a partially observed variable. We assessed their performance on the coefficients for (i) the interaction term, (ii) the partially observed variable, and (iii) the fully observed variable in the interaction. SMCFCS imputation consistently provided the least biased estimate of the three coefficients, except when there was a low prevalence of the fully observed variable in which case Just Another Variable (JAV) and passive imputation approaches showed the least bias. The Stratify-Impute-Append (SIA) approach also provided the least biased estimate for the three coefficients of interest, except when the fully observed variable had an underlying continuous form. \\
    
    SMCFCS showed good or optimal coverage for all scenarios. SIA performed similarly to SMCFCS, except when the fully observed variable had an underlying continuous form. JAV consistently showed large undercoverage for the coefficients of the fully observed variable and the interaction. This is most likely driven by the previously noted large bias, which leads to the severe type II error (\cite{Burton2006TheStatistics}). For the relative error, SMCFCS consistently provided a close approximation between the average model-based standard error and the empirical standard error, except when there was a low prevalence of the fully observed variable. In this case, SIA approach had the smallest relative error of the four imputation approaches but this was still very large. \\
    
    Careful consideration should be given to which parameter is of interest to the study when using JAV imputation. Although all four imputation approaches have similar bias when estimating the coefficient of the partially observed variable, JAV imputation will not only be biased for the interaction term but will also introduce bias into the coefficient for the fully observed variable that is included in the interaction. If the coefficient of the partially observed variable is the primary parameter of interest and the interaction term (and the coefficient of the fully observed variable) is a nuisance parameter, then any of the four methods are appropriate. However, if all three parameters are of interest, as is most often the case, SIA and SMCFCS should be used because they will give the least biased estimates of the slope parameters for $X$, $Z$, and $XZ$. \\
    
    
    Seaman \textit{et al} (2012) had previously showed that JAV imputation should not be used for logistic regression models with a \textit{quadratic term} (\cite{Seaman2012MultipleMethods}). Our results are consistent with their conclusions and extend the implications to the case of logistic regression models that include an interaction term. Passive imputation performed better than JAV in our settings, except when the fully observed variable in the interaction had an underlying continuous form, but we have shown that even in the simplest of specifications for a logistic regression model with interactions, both passive or JAV imputation should best be avoided. The only exception was that JAV performed almost as well as SMCFCS when the fully observed variable in the interaction has an underlying continuous form. However, this is a rare and specific example, and other imputation approaches may, more often, perform better. \\
    
    SMCFCS is possibly the best option for most analyses but SIA is also a suitable alternative, particularly in situations where the model being fit is not available in the SMCFCS package (e.g., excess hazard models in survival analysis). SIA allows one to remove the complexity of the interaction term and impute the missing data within levels of the categorical variable. SIA can be used when the fully observed variable is a categorical variable but only when there is a large enough number of observations for each level (i.e., avoiding perfect prediction). However, SIA is difficult to use if the fully observed variable in the interaction has a continuous underlying form and one would need to carefully consider how to categorise the continuous variable so that it appropriately captures the patterns observed in the variable; an approach could incorporate machine learning techniques such as classification and regression trees. Moreover, SIA cannot be used when both variables in the interaction contain missing values but SMCFCS can be used (passive and JAV can also be used but might provide biased results). \\
    
    We considered the simple scenario where the only complex covariate was an interaction. Seaman \textit{et al} (2012) investigated the performance of imputation approaches for logistic regression that included a quadratic term as the only complex covariate (\cite{Seaman2012MultipleMethods}). They found that, under MAR assumption, passive imputation produced biased estimates of the quadratic term and predictive mean matching (PMM) was unbiased with correct coverage. Since the SIA approach performs imputation within strata of the fully observed variable (i.e., removing the interaction and imputing only the partially observed covariate), it is possible that SIA can be used for more complex substantive models. For example, a logistic regression model with both a quadratic term ($X^{2}$ and an interaction ($XZ$) could be handled by using predictive mean matching within strata of the fully observed variable of the interaction. The logistic regression of the form $ Y = \beta_{0} + \beta_{1}Z + \beta_{2}C + \beta_{3}C^{2} + \beta_{4}X + \beta_{5}XZ $, where $C$ is a continuous variable with a quadratic effect on $Y$, then $X$ and $C$ could be imputed using PMM within strata of $Z$, yielding unbiased estimates for a complex logistic regression model. This hybrid approach would combine PMM and SIA, which we refer to as "stratified predictive mean matching". The performance of this hybrid approach, and in comparison to SMCFCS, requires further investigation. \\

    SMCFCS performed poorly for the bias and coverage of the coefficient of the fully observed variable (i.e., $Z_{5}$) in the interaction when there was a low prevalence (i.e., 1\%) of this fully observed variable. In this scenario (i.e., DGM 6), the average number of observations for which  $Z_{5}=1$ was 100, and the average number of observations for which $Z_{5}=0$ was 9900, in the simulated study. The low prevalence of this variable could lead to perfect prediction for the interaction term if there are low numbers for those with both $Z_{5}=1$ and $X_{1}=1$. Since the prevalence of $X_{1}=1$ was simulated to be 40\%, the approximate number of observations for $Z_{5}=1$ and $X_{1}=1$ in this scenario would be 40. SMCFCS could be adapted in this scenario with a data augmentation scheme where a few additional carefully crafted observations are added before the imputation model is fitted (\cite{White2010AvoidingVariables}). Another approach could be to use Firth's bias correction approach, which modifies the maximum likelihood procedure by removing its first order finite sample bias (\cite{Firth1993BiasEstimates,Kosmidis2021Jeffreys-priorModels}). We also simulated 100,000 observations to specifically assess the performance of SMCFCS in a larger sample, we found SMCFCS performed better with larger samples (results not shown). Further studies are required to assess the performance of these adaptations in various scenarios, particularly for low prevalence of the fully, or partially, observed variables. \\ 
    
    We compared the results of the four imputation methods using a real-world cancer data set on patients diagnosed with diffuse large B-cell lymphoma in England. As expected, SIA and SMCFCS showed very similar estimates for the effect (and confidence intervals) of diagnostic stage on mortality with 90 days since diagnosis. Passive imputation produced similar results to the complete case analysis, SIA, and SMCFCS approaches; however, as shown in the simulations, passive imputation is expected to be biased for the interaction term and the coefficient of the fully observed variable. In the simulations, JAV imputation showed markedly different (and poorer) results for the interaction term and the coefficient of the fully observed variable, which was also reflected in the cancer data analysis and would explain the large bias (away from the null) for these two parameters. \\

    Stage at diagnosis is often considered a categorical variable (i.e., stages I, II, III, and IV). We used a binary form for diagnostic stage (i.e., early vs late stage) for two reasons. Firstly, treatment, such as combination immunochemotherapy, for patients with diffuse large B-cell lymphoma is allocated based on an international prognostic index (IPI) (\cite{Project1993}). One criteria of the DLBCL-IPI is whether the cancer is late stage (i.e., stage III or IV) or early stage (i.e., stage I or II). Secondly, for simplicity of the simulation study, a categorical form for diagnostic stage would be handled similarly to a binary variable (i.e., impute within levels of the fully observed variable: comorbidity) and we anticipate this would not change the conclusions of our results. \\

    We assumed that missing data in diagnostic stage was \textit{missing at random}. In real-world settings, over the past 5 years, the proportion of missing diagnostic stage has reduced. The proportion of patients with a late diagnostic stage has either plateaued or increased at the same rate as the decrease in missing diagnostic stage, possibly indicating the two are related. Thus, the probability of observing cancer stage at diagnosis could be associated with the actual value of the cancer stage. This could potentially be true even after conditioning on all the observed covariates. Thus, the missing cancer diagnostic stage in our real-world study could be \textit{missing not at random} (MNAR), requiring alternative approaches to multiple imputation. Since the MNAR assumption is never testable with observed data, the standard approach is to perform sensitivity analyses to different missingness mechanisms (\cite{Gachau2020HandlingAssumption,Carpenter2013MultipleApplication}). \\

    SIA and SMCFCS imputation approaches give consistent estimation for logistic regression models with an interaction term when data are MAR, and either can be used in most analyses. SMCFCS performed better than SIA when the fully observed variable in the interaction has an underlying continuous form. Passive imputation performed better than JAV imputation but these approaches should only be used (instead of SMCFCS or SIA) when the coefficient of interest is the fully observed variable that has a low prevalence.

\newpage

\subsection*{Author contributions}
MJS, ENN and MQ developed the concept. MJS and MQ designed the first draft of the article and the computing code, interpreted and reviewed the code and the data, drafted and revised the article. All authors read and approved the final version of the article. MJS is the guarantor of the article.

\subsection*{Funding}

This research was funded by Cancer Research UK (Reference C7923/A18525). Funders had no role in the study design, data collection, data analysis, data interpretation, or writing of the report.

\subsection*{Institutional Review Board Statement}

We obtained the statutory approvals required for this research from the Confidentiality Advisory Group (CAG) of the Health Research Authority (HRA): PIAG 1–05(c) 2007. Ethical approval was obtained from the Research Ethics Committee (REC) of the Health Research Authority (HRA): 07/MRE01/52.

\subsection*{Informed Consent Statement}

Informed consent from participants was waived by the ethics committee. This work uses the data provided by patients and collected by the National Health Service as part of their care and support. We used anonymized National Cancer Registry and Hospital Episode Statistics data. No consent to participate was sought from patients. All methods were carried out in accordance with relevant guidelines and regulations.

\subsection*{Data availability Statement}

The data that support the findings of this study are available via application to the Public Health England Office for Data Release, but restrictions apply to the availability of these data.

\subsection*{Conflict of interest}

The authors declare no potential conflict of interests.

\newpage

\bibliographystyle{unsrtnat}
\bibliography{references}  






\newpage

\appendix

\section*{Appendix}\label{appendix1}

    \begin{sidewaystable}
        \resizebox{23cm}{!}{
        \begin{threeparttable}[ht]
        \centering
        \caption{Performance measures (relative bias, coverage, and relative error) of each imputation method in estimating the coefficients (fully observed, partially observed, and interaction), by data-generating mechanism.}
        \label{tab:TableA1}
        \begin{tabular}{lrlrrrrrrrrrrrrrrrrrrrr}
        \textbf{} & \textit{} &  & \multicolumn{6}{c}{\textbf{Fully observed (Z)}} & \multicolumn{1}{c}{\textbf{}} & \multicolumn{6}{c}{\textbf{Partially observed (X)}} & \multicolumn{1}{c}{\textbf{}} & \multicolumn{6}{c}{\textbf{Interaction (XZ)}} \\ \cline{4-23} 
        \textbf{} & \textit{} &  & \multicolumn{1}{c}{\textbf{\begin{tabular}[c]{@{}c@{}}R. Bias\\ (\%)\end{tabular}}} & \multicolumn{1}{c}{\textbf{MCSE}} & \multicolumn{1}{c}{\textbf{Cov.}} & \multicolumn{1}{c}{\textbf{MCSE}} & \multicolumn{1}{c}{\textbf{R. Error}} & \multicolumn{1}{c}{\textbf{MCSE}} & \multicolumn{1}{c}{\textbf{}} & \multicolumn{1}{c}{\textbf{\begin{tabular}[c]{@{}c@{}}R. Bias\\ (\%)\end{tabular}}} & \multicolumn{1}{c}{\textbf{MCSE}} & \multicolumn{1}{c}{\textbf{Cov.}} & \multicolumn{1}{c}{\textbf{MCSE}} & \multicolumn{1}{c}{\textbf{R. Error}} & \multicolumn{1}{c}{\textbf{MCSE}} & \multicolumn{1}{c}{\textbf{}} & \multicolumn{1}{c}{\textbf{\begin{tabular}[c]{@{}c@{}}R. Bias\\ (\%)\end{tabular}}} & \multicolumn{1}{c}{\textbf{MCSE}} & \multicolumn{1}{c}{\textbf{Cov.}} & \multicolumn{1}{c}{\textbf{MCSE}} & \multicolumn{1}{c}{\textbf{R. Error}} & \multicolumn{1}{c}{\textbf{MCSE}} \\ \cline{4-9} \cline{11-16} \cline{18-23} 
        \multicolumn{2}{l}{\textbf{No interaction}} &  & \multicolumn{1}{l}{} & \multicolumn{1}{l}{} & \multicolumn{1}{l}{} & \multicolumn{1}{l}{} & \multicolumn{1}{l}{} & \multicolumn{1}{l}{} & \multicolumn{1}{l}{} & \multicolumn{1}{l}{} & \multicolumn{1}{l}{} & \multicolumn{1}{l}{} & \multicolumn{1}{l}{} & \multicolumn{1}{l}{} & \multicolumn{1}{l}{} & \multicolumn{1}{l}{} & \multicolumn{1}{l}{} & \multicolumn{1}{l}{} & \multicolumn{1}{l}{} & \multicolumn{1}{l}{} & \multicolumn{1}{l}{} & \multicolumn{1}{l}{} \\
        \textbf{} & \textit{Passive} &  & 0.0 & 0.003 & 93.4 & 0.008 & -4.0 & 2.149 &  & 0.0 & 0.004 & 94.7 & 0.007 & 1.7 & 2.294 &  & - & - & - & - & - & - \\
        \textbf{} & \textit{JAV} &  & 0.0 & 0.003 & 93.4 & 0.008 & -4.0 & 2.149 &  & 0.0 & 0.004 & 94.7 & 0.007 & 1.7 & 2.294 &  & - & - & - & - & - & - \\
        \textbf{} & \textit{SIA} &  & 0.0 & 0.003 & 93.7 & 0.008 & -4.0 & 2.149 &  & 0.0 & 0.004 & 94.9 & 0.007 & 1.5 & 2.289 &  & - & - & - & - & - & - \\
        \textbf{} & \textit{SMCFCS} &  & 0.0 & 0.003 & 93.6 & 0.008 & -4.0 & 2.149 &  & 0.0 & 0.004 & 95.1 & 0.007 & 0.9 & 2.276 &  & - & - & - & - & - & - \\
        \multicolumn{2}{l}{\textbf{DGM1}} &  &  &  &  &  &  &  &  &  &  &  &  &  &  &  &  &  &  &  &  &  \\
        \textbf{} & \textit{Passive} &  & -50.7 & 0.004 & 95.6 & 0.007 & 24.4 & 2.857 &  & 4.5 & 0.004 & 96.0 & 0.006 & 6.1 & 2.440 &  & -44.3 & 0.00 & 95.9 & 0.006 & 61.1 & 2.993 \\
        \textbf{} & \textit{JAV} &  & -162.4 & 0.004 & 75.4 & 0.014 & 22.8 & 2.960 &  & -3.7 & 0.004 & 95.5 & 0.007 & 3.7 & 2.421 &  & -119.5 & 0.01 & 81.8 & 0.012 & 52.5 & 4.277 \\
        \textbf{} & \textit{SIA} &  & 8.1 & 0.005 & 95.2 & 0.007 & 0.4 & 2.324 &  & -0.5 & 0.004 & 94.9 & 0.007 & -0.3 & 2.297 &  & -1.1 & 0.01 & 94.1 & 0.008 & 0.7 & 1.713 \\
        \textbf{} & \textit{SMCFCS} &  & 6.0 & 0.005 & 95.1 & 0.007 & 1.2 & 2.338 &  & -0.9 & 0.004 & 94.6 & 0.007 & 0.4 & 2.310 &  & -2.5 & 0.01 & 94.0 & 0.008 & 2.1 & 1.724 \\
        \multicolumn{2}{l}{\textbf{DGM2}} &  &  &  &  &  &  &  &  &  &  &  &  &  &  &  &  &  &  &  &  &  \\
        \textbf{} & \textit{Passive} &  & -13.4 & 0.004 & 98.2 & 0.004 & 23.5 & 2.837 &  & 0.9 & 0.004 & 96.6 & 0.006 & 6.0 & 2.440 &  & -118.3 & 0.00 & 99.4 & 0.002 & 57.5 & 3.413 \\
        \textbf{} & \textit{JAV} &  & -109.8 & 0.004 & 82.8 & 0.012 & 22.0 & 2.985 &  & -3.6 & 0.004 & 96.3 & 0.006 & 3.6 & 2.426 &  & -478.4 & 0.01 & 86.4 & 0.011 & 53.7 & 5.181 \\
        \textbf{} & \textit{SIA} &  & 8.5 & 0.005 & 95.8 & 0.006 & 0.2 & 2.319 &  & -0.7 & 0.004 & 94.6 & 0.007 & -1.2 & 2.274 &  & 454.2 & 0.01 & 94.9 & 0.007 & -0.4 & 2.072 \\
        \textbf{} & \textit{SMCFCS} &  & 7.2 & 0.005 & 95.1 & 0.007 & 0.8 & 2.330 &  & -0.6 & 0.004 & 94.6 & 0.007 & -0.4 & 2.292 &  & 100.6 & 0.01 & 94.7 & 0.007 & 0.8 & 2.074 \\
        \multicolumn{2}{l}{\textbf{DGM3}} &  &  &  &  &  &  &  &  &  &  &  &  &  &  &  &  &  &  &  &  &  \\
        \textbf{} & \textit{Passive} &  & -118.0 & 0.004 & 83.7 & 0.012 & 25.9 & 2.892 &  & 11.6 & 0.004 & 93.6 & 0.008 & 4.1 & 2.395 &  & -103.3 & 0.00 & 83.4 & 0.006 & 56.4 & 2.582 \\
        \textbf{} & \textit{JAV} &  & -249.3 & 0.004 & 56.2 & 0.016 & 24.0 & 3.062 &  & -4.0 & 0.004 & 95.8 & 0.006 & 2.4 & 2.427 &  & -168.0 & 0.01 & 67.7 & 0.008 & 46.8 & 3.729 \\
        \textbf{} & \textit{SIA} &  & 6.2 & 0.005 & 95.3 & 0.007 & 0.4 & 2.323 &  & -0.9 & 0.004 & 94.5 & 0.007 & -2.0 & 2.257 &  & -0.5 & 0.01 & 95.5 & 0.006 & 1.3 & 1.313 \\
        \textbf{} & \textit{SMCFCS} &  & 6.6 & 0.005 & 96.0 & 0.006 & 0.1 & 2.316 &  & -1.5 & 0.004 & 94.6 & 0.007 & -2.1 & 2.253 &  & -64.0 & 0.01 & 94.9 & 0.007 & 0.3 & 1.339 \\
        \multicolumn{2}{l}{\textbf{DGM4}} &  &  &  &  &  &  &  &  &  &  &  &  &  &  &  &  &  &  &  &  &  \\
        \textbf{} & \textit{Passive} &  & -23.1 & 0.004 & 95.1 & 0.007 & 9.9 & 2.519 &  & 4.2 & 0.003 & 96.6 & 0.006 & 5.8 & 2.426 &  & -22.5 & 0.01 & 96.8 & 0.006 & 21.6 & 2.193 \\
        \textbf{} & \textit{JAV} &  & -57.6 & 0.004 & 87.9 & 0.010 & 2.5 & 2.354 &  & -1.3 & 0.004 & 96.2 & 0.007 & 1.0 & 2.317 &  & -24.8 & 0.01 & 92.6 & 0.008 & 8.2 & 2.010 \\
        \textbf{} & \textit{SIA} &  & 2.5 & 0.004 & 94.8 & 0.007 & 0.0 & 2.294 &  & -0.6 & 0.004 & 95.2 & 0.007 & 1.0 & 2.317 &  & -2.7 & 0.01 & 95.8 & 0.006 & 0.5 & 1.699 \\
        \textbf{} & \textit{SMCFCS} &  & 2.8 & 0.004 & 95.1 & 0.007 & -0.1 & 2.290 &  & -0.6 & 0.004 & 95.3 & 0.007 & -0.5 & 2.282 &  & -2.4 & 0.01 & 95.3 & 0.007 & 0.3 & 1.702 \\
        \multicolumn{2}{l}{\textbf{DGM5}} &  &  &  &  &  &  &  &  &  &  &  &  &  &  &  &  &  &  &  &  &  \\
        \textbf{} & \textit{Passive} &  & -46.8 & 0.003 & 95.6 & 0.007 & 24.6 & 2.856 &  & -6.7 & 0.003 & 94.5 & 0.007 & 1.7 & 2.348 &  & -38.3 & 0.00 & 95.9 & 0.006 & 46.6 & 2.785 \\
        \textbf{} & \textit{JAV} &  & -22.6 & 0.004 & 94.7 & 0.007 & 4.4 & 2.394 &  & -3.7 & 0.003 & 94.3 & 0.007 & 1.5 & 2.342 &  & -13.8 & 0.01 & 93.7 & 0.008 & -1.8 & 1.752 \\
        \textbf{} & \textit{SIA} &  & -68.3 & 0.001 & 62.3 & 0.015 & 6.1 & 2.437 &  & -77.1 & 0.008 & 75.3 & 0.014 & 1.2 & 2.329 &  & -68.0 & 0.00 & 35.6 & 0.015 & 1.9 & 2.021 \\
        \textbf{} & \textit{SMCFCS} &  & -1.8 & 0.004 & 95.1 & 0.007 & 4.4 & 2.398 &  & -1.6 & 0.003 & 94.8 & 0.007 & 1.9 & 2.351 &  & -1.2 & 0.01 & 94.6 & 0.007 & -0.2 & 1.691 \\
        \multicolumn{2}{l}{\textbf{DGM6}} &  &  &  &  &  &  &  &  &  &  &  &  &  &  &  &  &  &  &  &  &  \\
        \textbf{} & \textit{Passive} &  & 662.5 & 0.067 & 98.1 & 0.004 & 2329.8 & 140.363 &  & -0.3 & 0.003 & 94.4 & 0.007 & 0.6 & 2.320 &  & -106.6 & 0.09 & 100.0 & 0.000 & 2757.1 & 3.477 \\
        \textbf{} & \textit{JAV} &  & 638.7 & 0.052 & 99.4 & 0.002 & 2406.7 & 173.719 &  & -1.5 & 0.003 & 95.0 & 0.007 & 1.0 & 2.333 &  & -237.5 & 0.09 & 100.0 & 0.000 & 2949.0 & 4.877 \\
        \textbf{} & \textit{SIA} &  & 1925.6 & 0.123 & 97.1 & 0.005 & 2098.4 & 85.316 &  & -0.8 & 0.003 & 95.0 & 0.007 & 1.2 & 2.334 &  & 101.3 & 0.18 & 99.5 & 0.002 & 1995.8 & 1.496 \\
        \textbf{} & \textit{SMCFCS} &  & 1298.1 & 0.085 & 97.8 & 0.005 & 2562.1 & 110.709 &  & -0.6 & 0.003 & 95.1 & 0.007 & -0.1 & 2.305 &  & -27.1 & 0.12 & 99.5 & 0.002 & 2774.8 & 1.510 \\ \hline
        \end{tabular}
        \begin{tablenotes}
            \item \textbf{R. Bias}: relative bias, \textbf{Cov.}: coverage of 95\% confidence intervals, \textbf{R. Error}: relative error of model-based standard error and empirical error, \textbf{MCSE}: Monte-Carlo Standard Error, \textbf{JAV}: Just Another Variable, \textbf{SIA}: Stratify-Impute-Append, \textbf{SMCFCS}: substantive model compatible fully conditional specification, \textbf{DGM}: data-generating mechanism.
        \end{tablenotes}
        \end{threeparttable}}
    \end{sidewaystable}

\end{document}